\documentclass[a4paper,12pt]{article}

\usepackage{amsmath,amssymb,epsfig}

\numberwithin{equation}{section}       

\newcommand{\dd}{\mathrm{d}}
\newcommand{\ee}{\mathrm{e}}

\newcommand{\bbZ}{\mathbb{Z}}

\newcommand{\bbC}{\mathbb{C}}
\newcommand{\bbP}{\mathbb{P}}
\newcommand{\bbF}{\mathbb{F}}

\newcommand{\rep}[1]{\mathbf{#1}}

\DeclareMathOperator{\SU}{\mathit{SU}}
\DeclareMathOperator{\SO}{\mathit{SO}}
\DeclareMathOperator{\Spin}{\mathit{Spin}}
\DeclareMathOperator{\so}{\mathit{so}}
\DeclareMathOperator{\su}{\mathit{su}}
\DeclareMathOperator{\symp}{\mathit{sp}}

\DeclareMathOperator{\tr}{tr}
\DeclareMathOperator{\diag}{diag}

\DeclareMathOperator{\cc}{c}

\DeclareMathOperator{\KI}{\mathit{I}}
\DeclareMathOperator{\KII}{\mathit{II}}
\DeclareMathOperator{\KIII}{\mathit{III}}
\DeclareMathOperator{\KIV}{\mathit{IV}}

\newcommand{\beqn}{\begin{eqnarray}}
\newcommand{\eeqn}{\end{eqnarray}}
\def\be#1\ee{\begin{equation}#1\end{equation}}

\begin{document}


\begin{titlepage}

\vfill

\begin{flushright}
QMUL-PH-02-08\\
hep-th/0204228\\
\end{flushright}

\vfill

\begin{center}
\begin{LARGE}
F-theory duals of M-theory on $S^1/\bbZ_2\times T^4/\bbZ_N$
\bigskip
\end{LARGE}
\end{center}

\begin{center}
Monika Marquart$^{1,2}$ and Daniel Waldram$^{2,3}$
\end{center}

\begin{center}
\begin{small}
   $^1$\textit{Fachbereich Physik, Martin-Luther-Universit\"{a}t
   Halle-Wittenberg\\
   Friedemann-Bach-Platz 6,\\
   D-06099 Halle, Germany}\\
   \bigskip
   $^2$\textit{Department of Physics\\
   Queen Mary, University of London\\
   Mile End Rd, London E1 4NS, UK}\\
   \bigskip
   $^3$\textit{Isaac Newton Institute for Mathematical Sciences\\
   University of Cambridge\\
   20 Clarkson Road, Cambridge, CB3 0EH, U.K.}\\
\end{small}
\end{center}

\bigskip

\begin{center}
\textbf{Abstract}
\end{center}

\begin{quote}
In this note, we use results of Aspinwall and Morrison to discuss the
F-theory duals of certain $T^4/\bbZ_N$ orbifold compactifications of
Ho\v{r}ava--Witten theory. In the M-theory limit an interesting
set of rules, based on anomaly cancellation, has been developed for
what gauge and matter multiplets must be present on the various
orbifold fixed planes. Here we show how several aspects of these rules
can be understood directly from F-theory. 
\end{quote}

\vfill

\end{titlepage}


\section{Introduction}
\label{sec:intro}

The description of orbifold compactifications of M-theory began with
the seminal paper of Ho\v{r}ava and Witten~\cite{HW1} which realised the
strongly coupled $E_8\times E_8$ heterotic string as
eleven-dimensional M-theory compactified on $S^1/\bbZ_2$. 
Subsequently, a number of authors have considered other orbifold
compactifications. Models based on orbifolds of $T^5$ were considered
in~\cite{WI1} and~\cite{DM,GM} (see also~\cite{MO}). More recently,
two groups~\cite{FLO,KSTY} have made a general analysis of
Ho\v{r}ava--Witten type compactifications on spaces of the form
$S^1/\bbZ_2\times Y$, where $Y=T^4/\bbZ_N$ is an orbifold limit of
K3. These models preserve $\mathcal{N}=1$ supersymmetry in six
dimensions, and have a weak-coupling limit given by the heterotic
string compactified on $Y$.  

Since there is no fundamental formulation of M-theory, in each case,
following~\cite{HW1}, it has been necessary to use consistency
arguments, in particular the requirements of anomaly cancellation, to
deduce what twisted matter and interactions reside on the orbifold
fixed planes. This has led particularly the authors of~\cite{FLO} and
also of~\cite{KSTY} to derive a very interesting and complete set of
rules for constucting such models. These are, in general, far from
intuitive from a geometrical perspective. One way to understand this
is that there are necessarily large M-theory corrections to the
geometry near the orbifold fixed planes. The rules are of particular
interest because they can be generalised to describe supersymmetric
orbifold compactifications to four-dimensions~\cite{DFO2} which may
provide interesting new avenues for model-building.

Recently, it has been possible to justify the rules developed
in~\cite{FLO,KSTY} by considering duality to type I' superstring
compactifications~\cite{GKSTY}. Here, we take an alternative approach,
using results of Aspinwall and Morrison~\cite{AM1,AM2} to describe
the compactifications by their F-theory duals. These results are not
new, but by taking the M-theory limit with $Y$ large~\cite{A,AM1} one
gets a simple understanding of several of the rules developed
in~\cite{FLO,KSTY}. (An obvious complementary approach would
be to use the toric description of the F-theory models as for instance
in~\cite{toric}.) We concentrate on the example where $Y=T^4/\bbZ_2$
and the $E_8$ gauge bundles are such that a perturbative
$\SO(16)\times[\SU(2)\times E_7]$ subgroup is preserved. This model
has the interesting property of matter charged under groups from both
$E_8$ ten-planes. In the final section, we discuss
generalisations to F-theory limits of compactifications with various
other bundles and orbifolds. 


\section{Ho\v{r}ava--Witten theory on $Y=T^4/\bbZ_N$}
\label{sec:HWreview}

We start by reviewing the description of orbifold compactifications
of Ho\v{r}ava--Witten theory on $Y$ following the work of~\cite{FLO}
and~\cite{KSTY}. For definiteness, we will concentrate on the duals of
perturbative heterotic models on $Y=T^4/\bbZ_2$, though at the end of
the paper we will briefly consider other orbifold limits of K3 and
gauge groups. 

\subsection{Heterotic string limit, orbifolds and fractional
  instantons} 
\label{sec:het}

First, consider the $E_8\times E_8$ heterotic string compactified on
$Y$. If $Y=T^4/\bbZ_2$, the orbifold has sixteen fixed points, each
giving an $A_1$ singularity where the geometry is locally
$\bbC^2/\bbZ_2$. For the gauge fields in perturbative backgrounds,
one typically assumes that the gauge bundle on the unquotiented space
$T^4$ is flat and then makes some identification of the gauge bundle
fibres under the $\bbZ_2$ orbifold action. The bundle on $Y$ then
has, at most, $\bbZ_2$ holonomy around the $A_1$ singularities. Any
field strength is localized at the singularity and if the holonomy is
non-trivial, this gives a ``fractional instanton''~\cite{B}, with
possibly non-integral instanton charge.  

There are basically three possibilities for embedding the
$\bbZ_2$ holonomy in $E_8$: it can be trivial, in which case the
unbroken gauge symmetry remains $E_8$, it can break $E_8$ to
$\Spin(16)/\bbZ_2$, or it can break $E_8$ to $[\SU(2)\times
E_7]/\bbZ_2$. (These latter two we will refer to loosely as $\SO(16)$
and as $\SU(2)\times E_7$.) 

For perturbative backgrounds two possibilities
arise~\cite{W,E}. Either one $E_8$ is trivial while the other is
broken so the final symmetry is $E_8\times[\SU(2)\times E_7]$, or both
are broken giving $\SO(16)\times [\SU(2)\times E_7]$. The former case
is the so called ``standard embedding''. In this note, we will
concentrate on the latter case. One can calculate the fractional
instanton charge at each $A_1$ singularity associated with the given
$\bbZ_2$-bundle. One finds that for the latter case, each fractional
instanton giving the $\SO(16)$ factor has charge one. For the
$\SU(2)\times E_7$ factor, on the other hand, each fractional
instanton has charge $1/2$. For the standard embedding the
corresponding instantons also have charge $3/2$. Thus in both cases
the net charge per singularity is $3/2$. 

An important requirement of heterotic compactifications is that, as a
result of anomaly cancellation, the net ``magnetic charge'', as given
by $\cc_2(V_1)+\cc_2(V_2)-\cc_2(TY)$, must vanish. Here
$\cc_2(V_1)+\cc_2(V_2)$ is the total gauge instanton charge given by
the sum of the second Chern classes of the two $E_8$ gauge bundles $V_1$
and $V_2$. The last term is the second Chern class of the tangent
bundle of the compact manifold $Y$, with $\cc_2(TY)=24$ for a K3
surface.  Thus if $Y$ is a $T^4/\bbZ_2$ orbifold, the last term gives
a contribution of $-24/16=-3/2$ for each $A_1$ singularity. 
Consequently the net gauge instanton number per singularity must be
$3/2$ as is indeed the case for the two perturbative examples
discussed above.     

The full spectrum of the theory is given in~\cite{W,E}. For our main
example the unbroken gauge group is $\SO(16)\times[\SU(2)\times E_7]$.
Given $\mathcal{N}=1$ supersymmetry in six dimensions, we can expect
matter in hypermultiplets or tensor multiplets as well as the gauge
vector multiplets. The breaking $E_8\to\SO(16)$ leads to a
decomposition $\rep{248}\to(\rep{120}+\rep{128})$. This gives the
$120$ vector multiplets transforming in the adjoint representation and
$128$ hypermultiplets transforming in the spinor representation of
$SO(16)$. Meanwhile the breaking $E_8\to \SU(2)\times E_7$ leads to a
decomposition
$(\bf{248})\to(\rep{1},\rep{133})+(\rep{3},\rep{1})+(\rep{2},\rep{56})$.
The first two factors give the $133$ vector multiplets in the adjoint
representation of $E_7$ and three vector multiplets in the adjoint
representation of $SU(2)$. The third factor gives $112$
hypermultiplets in the bifundamental representation. In addition the
compactification leads to four moduli which are hypermultiplets and
gauge singlets and there is one universal tensor multiplet which
includes the dilaton. Thus far this is just the untwisted massless
spectrum of the weakly coupled heterotic theory. In addition there are
twisted string states on the $T^4/\mathbb{Z}_2$ orbifold. These give
sixteen half hypermultiplets transforming as
$\frac{1}{2}(\rep{16},\rep{2},\rep{1})$ and so are charged under
\textit{both} factors in the perturbative gauge group
$E_8\times[\SU(2)\times E_7]$. 

\subsection{Ho\v{r}ava--Witten geometry}
\label{sec:HWlimit}

Ho\v{r}ava--Witten theory gives the strong coupling limit of the
$E_8\times E_8$ heterotic string as M-theory compactified on
$S^1/\bbZ_2$. Thus here our starting point is M-theory compactified on
the product of two $\bbZ_2$ orbifolds $T^4/\mathbb{Z}_2\times
S^1/\mathbb{Z}_2$. The orbifold projection acting on $S^1$ leaves two
fixed ten-planes. Following Ho\v{r}ava and Witten, there are $E_8$
vector multiplets localized on each fixed ten plane. The action of the
$\mathbb{Z}_2$ on $T^4$ leaves sixteen fixed seven-planes. The
combined action of both orbifolds results in sixteen pairs of fixed
six-planes, which are the intersections of the two fixed ten-planes
and the sixteen fixed seven-planes.  

In general, one expects additional gauge and matter degrees of freedom
on the fixed seven- and six-planes. The compactification of M-theory
on $T^4/\bbZ_2$ gives a theory with sixteen supercharges so that the
only possible multiplets on the fixed seven-planes are
seven-dimensional vectors. Compactifying further on $S^1/\bbZ_2$
breaks half of the supersymmetry to eight supercharges. Thus we can
have six-dimensional hypermultiplets or vector multiplets on the fixed
six-planes. 

A priori, since there is no fundamental formulation of M-theory it is
unclear what new multiplets appear. However, following Ho\v{r}ava and
Witten, the new degrees of freedom can be deduced from the requirement
of anomaly cancellation~\cite{FLO,KSTY}. On the six-dimensional planes
this is a particularly powerful tool as there are gauge, gravitational
and mixed anomalies. Since there are no chiral anomalies in odd
dimensions, it would appear to be impossible to determine the matter
content on the seven-planes. However, by considering those parts of the
seven-dimensional multiplets which survive the $\bbZ_2$ projection
onto the six-planes, in turns out that the six-dimensional anomalies
are basically sufficient to determine the seven-dimensional content. 

These kinds of arguments have led to a set of local rules for what
matter and gauge groups are present on each fixed plane for different
$T^4/\bbZ_N$ orbifolds~\cite{FLO,KSTY}. In particular, one finds the
following for the perturbative $\SO(16)\times[\SU(2)\times E_7]$
model. First, one assumes the $E_8$ bundles on the two fixed
ten-planes have the same $\bbZ_2$ holonomies as in the weakly coupled
string limit, so again there are  fractional instantons localised on the
six-planes, giving an unbroken  perturbative
$\SO(16)\times[\SU(2)\times E_7]$ gauge group. The familiar untwisted
states of the string theory limit then appear as zero-modes of the
$E_8$ vector multiplets in this background. Anomaly cancellation
implies that there must be $\SU(2)$ vector multiplets on each of the
fixed seven-planes. This is as expected given that these give planes
of $A_1$ singularities, so the geometrical blow-up modes together
with wrapped M2-brane states should form an $\SU(2)$ gauge
multiplet. Naively one would expect these to be new non-perturbative
degrees of freedom so the full low-energy gauge symmetry becomes  
\begin{equation}
   \SO(16) \times [\SU(2)\times E_7] 
       \times \SU(2)_{\text{non-pert}}^{16} .
\end{equation}

An important issue is how to identify the twisted string states which
are charged under both the perturbative $\SO(16)$ and $\SU(2)$
groups. The problem is that in Ho\v{r}ava--Witten theory these live on
different separated fixed ten-planes, and so it appears no localised
state could be charged under both factors. 

The minimal solution to this problem, consistent with anomaly
cancellation, was given in~\cite{FLO} and~\cite{KSTY}. The point is
that, on each of the six-planes where the sixteen seven-planes
intersect $\SU(2)\times E_7$ ten-plane, one must identity the
non-perturbative seven-dimensional $\SU(2)$ gauge fields with the
perturbative $\SU(2)$ fields. This correlates gauge
transformations on the two factors so in the language of~\cite{FLO}
there is simply a single $\SU(2)$ gauge group extending over one
ten-plane and the sixteen seven-planes. In~\cite{KSTY}, this
``locking'' of the gauge groups is characterised by saying that the
gauge factor visible in the heterotic string description is the
diagonal $\SU(2)^*=\diag(\SU(2)\times\SU(2)^{16}_{\text{non-pert}})$
of the  product of the non-perturbative and perturbative groups. Both
papers~\cite{FLO} and~\cite{KSTY} identify the low-energy 
six-dimensional gauge group as having a single $\SU(2)$ factor
\begin{equation}
   \SO(16) \times [\SU(2)^* \times E_7] .
\end{equation}
Here we will interpret this as implying that the zero modes for the
gauge fields of this single $\SU(2)^*$ factor extend over both one
fixed ten-plane and the sixteen fixed seven-planes. It is then argued
that the twisted matter lives on the other set of six-planes where the
seven-planes intersect the $\SO(16)$ ten-plane. The matter is charged
under the non-perturbative $\SU(2)$ factor, but because of the locking
this means it is, in fact, also charged under the single perturbative
$\SU(2)$ on the other ten-plane (essentially because the zero mode
extends over both the ten- and seven-planes). This explains how the
twisted matter can be both local and carry the correct charge.    

One other, counter-intuitive related rule is required in order to
cancel all the anomalies. Naively one expects the fields $\Phi_i$ of the
seven-dimensional vector multiplets to have definite transformation
properties under the $S^1/\bbZ_2$ orbifold projection: so that
$\Phi_i(x^{11})=\Phi_i(-x^{11})$ or $\Phi_i(x^{11})=-\Phi_i(-x^{11})$. In
general, the seven-dimensional vector multiplet splits into a vector
multiplet and a hypermultiplet in six-dimensions. Under the projection
one expects one or other of these multiplets to remain
massless. Remarkably the anomaly rules require that, in general, one
must allow for \textit{different} parts of the seven-dimensional
vector multiplet to survive the projection to each of the two
six-planes at the intersections of the fixed seven- and ten-planes.  

This is very unexpected, since, if the compact space is really
$T^4/\bbZ_2 \times S^1/\bbZ_2$, then we would expect the
seven-dimensional fields to have definite transformation properties
and the \textit{same} part of the multiplet would survive on each
six-plane. Similarly, it is hard to understand, purely
geometrically, why one should identify the perturbative and apparently
non-perturbative gauge groups as a single $\SU(2)^*$ factor. 

The solution is that the compact space cannot in fact be a
product. The presence of magnetic charges at each $A_1$ singularity
acts as a source which necessarily distorts the space. The point is
that in Ho\v{r}ava--Witten theory the gauge fields and Riemann
curvature on each ten-plane $M_1$ and $M_2$ couple magnetically to the
four-form $G$ of the bulk eleven-dimensional supergravity~\cite{HW1}
(as well as providing a source of stress-energy in Einstein's
equations~\cite{LOW}). One has  
\begin{multline}
   \dd G \sim 
       m_P^{-3}\left[\tr(F_1\wedge F_1) - \frac{1}{2}\tr(R\wedge R) \right] 
           \wedge \delta(M_1) \\
       + m_P^{-3}\left[\tr(F_2\wedge F_2) - \frac{1}{2}\tr(R\wedge R) \right] 
           \wedge \delta(M_2) ,
\label{eq:dG}
\end{multline}
where $F_1$ and $F_2$ are the two $E_8$ gauge field strengths and
$m_P$ is the eleven-dimensional Planck mass. Note that the integral of
the right-hand side of the equation over $S^1/\bbZ_2\times Y$ gives
the net magnetic charge $\cc_2(V_1)+\cc_2(V_1)-\cc_2(TY)$ which we know
from above vanishes, as it must  since $\dd G$ is exact. However,
the sources in general do not cancel at every point. For instance, in
the orbifold compactifications all the $\tr(F_i\wedge F_i)$ and
$\tr(R\wedge R)$ charge is localised at the orbifold singularities,
and is proportional to the corresponding contributions to the second
Chern classes.  These need not cancel at
each $A_1$ singularity on each ten-plane separately. In particular, we
see that there is a net charge on the $\SO(16)$ ten-plane of $-1/4$
per $A_1$ singularity, and a net charge on the $\SU(2)\times E_7$
ten-plane of $+1/4$ per $A_1$ singularity. As a result $G\neq 0$ in
the eleven-dimensional bulk and, as in~\cite{WW}, the manifold
deforms from a simple product. For smooth compactifications one can
suppress this effect by choosing $Y$ to be very large and the gauge
fields slowly varying so that the sources are small with respect
to $m_P$. However, in the orbifold limit, there is no scale to the
curvature of $Y$ or the gauge field strength since both are
singular and there is no analogous suppression.  

In summary, aside from~\cite{GKSTY}, there has been no derivation of the
M-theory rules~\cite{FLO,KSTY} from first principles. It is only
possible to show that the anomalies cancel using this
recipe. It may be possible to gain additional insight by considering the
full deformed M-theory geometry, in particular, how this could lead to
the identification of $\SU(2)$ factors and the projections on the
seven-dimensional multiplets. Here, instead, we will consider
the F-theory duals to justify the rules, though note this will also
provide additional evidence that the M-theory background is
deformed.


\section{F-theory description}
\label{sec:Ftheory}

In this section we consider the F-theory formulation of the
$SO(16)\times[E_7\times SU(2)]$ model. We will see that the matter
content and gauge groups can be derived directly. In particular, we
justify the identification of a heterotic $SU(2)^*$ gauge group and
the appearance of twisted matter states discussed at the end of the
last section. We should point out that these F-theory models are not
new but are a simple case of a class of models considered
in~\cite{AM2}. The results are briefly generalized to other models in
the next section.      

\subsection{F-theory and the stable degeneration limit}
\label{sec:Fintro}

Let us first summarize the duality in six dimensions between F-theory
compactified on a Calabi-Yau threefold $X$ and the heterotic string
compactified on an elliptically fibred K3 surface
$Y$~\cite{Ftheory}. To keep the problem as simple as possible we
restrict ourselves to describing the classical geometry of $Y$, which
means that both the base $\bbP^1_B$ and the elliptic fibre of $Y$
are large. This corresponds on the F-theory side to taking a
particular limit of the threefold known as a stable degeneration,
first discussed in \cite{FMW} and explained in detail in \cite{AM1}. 

Recall that under the duality, the elliptic fibers of the heterotic K3
manifold $Y$ are replaced by K3 fibers in the F-theory threefold
$X$. These fibers are themselves elliptically fibred so that $X$ can
be viewed as an elliptic fibration over a Hirzebruch surface 
$\pi:X\to\bbF_n$. The Hirzebruch surface itself is a $\bbP^1$
fibration over the common base $\bbP^1_B$, giving a projection
$\bbF_n\to\bbP^1_B$.   

The elliptic fibration $\pi:X\to\bbF_n$, which, by definition, also has
a section $\sigma:\bbF_n\to X$, can be described via a Weierstrass
model 
\begin{equation}
   y^2 = x^3 + a(s,t)x + b(s,t) ,
\label{weierst}
\end{equation}
where $s$ and $t$ parametrise the base $\bbP^1_B$ and the fibre
$\bbP^1$ of $\bbF_n$. The affine coordinates $x$ and $y$ are sections
of $\mathcal{L}^2$ and $\mathcal{L}^3$ respectively where
$\mathcal{L}$ is the co-normal bundle to the $\bbF_n$ section in
$X$. Similarly $a$ and $b$ are sections of $\mathcal{L}^4$ and
$\mathcal{L}^{6}$. Since $X$ is Calabi--Yau, the canonical bundle $K_X$
is trivial, so that, by adjunction, $\mathcal{L}=K_{\bbF_n}^{-1}$. The
torus degenerates whenever the discriminant $\delta=24b^2+4a^3$, which
is a section of $\mathcal{L}^{12}$, vanishes. These degenerations
characterise the enhanced gauge symmetries of the theory, following
the classical Kodaira classification. The gauge group at a given point
in the base is determined by the order of vanishing of the sections
$\delta$, $a$ and $b$ as summarized the familiar list given in
Table~\ref{tab:sing} taken from~\cite{AM1}.
\begin{table}[htbp]
   \begin{center}
   \begin{tabular}{|ccc|c|c|c|}
   \hline 
   $o(a)$ & $o(b)$ & $o(\delta)$ & Kodaira fibre 
      & singularity & gauge algebra \\
   \hline 
   $\ge 0$ & $\ge 0$ & 0 & $\KI_0$    & - & - \\
   0 & 0     & 1       & $\KI_1$    & - & - \\
   0 & 0     & $2n\ge 2$ & $\KI_{2n}$ & $A_{2n-1}$ 
          & $\su(2n)$ or $\symp(2n)$ \\
   0 & 0 & $2n+1\ge 3$ & $\KI_{2n+1}$ & $A_{2n}$ 
          & $\su(2n+1)$ or $\so(2n+1)$ \\
   $\ge 1$ & 1 & 2 & $\KII$ & - & - \\
   1 & $\ge 2$ & 3 & $\KIII$ & $A_1$ & $\su(2)$ \\
   $\ge 2$ & 2 & 4 & $\KIV$ & $A_2$ & $\su(3)$ or $\su(2)$ \\
   $\ge 2$ & $\ge 3$  & 6 & $\KI_0^*$ & $D_4$ 
          & $\so(8)$ or $\so(7)$ or $g_2$ \\
   2 & 3 & $n+6\ge 7$ & $\KI_n^*$ & $D_{n+4}$ 
          & $\so(2n+8)$ or $\so(2n+7)$ \\
   $\ge 3$ & 1 & 8 & $\KIV^*$ & $E_6$ & $e_6$ or $f_4$ \\
   3 & $\ge 5$ & 9 & $\KIII^*$ & $E_7$ & $e_7$ \\
   $\ge 4$ & 5 & 10 & $\KII^*$ & $E_8$ & $e_8$ \\
   $\ge 4$ & $\ge 6$ & $\ge 12$  & non-minimal &  &  \\
   \hline
   \end{tabular}
   \end{center}
   \caption{Orders of vanishing, fibres, singularities and gauge algebra}
   \label{tab:sing}
\end{table}

If we write $L$ for the class of divisors associated to the vanishing
of sections of $\mathcal{L}$, since $\mathcal{L}=K_{\bbF_n}^{-1}$ we
have
\begin{equation}
   L = 2C_0 + (n+2)f ,
\label{eq:Ldef}
\end{equation}
where $C_0$ is the class of the exceptional divisor on $\bbF_n$ and
$f$ the class of the $\bbP^1$ fibre of $\bbF_n\to\bbP^1_B$. These form
a basis of divisor classes on $\bbF_n$ and have intersections
$C_0\cdot C_0=-n$, $C_0\cdot f=1$ and $f\cdot f=0$. The discriminant
curve $\Delta$ defined by $\delta=0$ is in the class $\Delta=12L$. 

\begin{figure}[htbp]
   \centerline{\epsfig{file=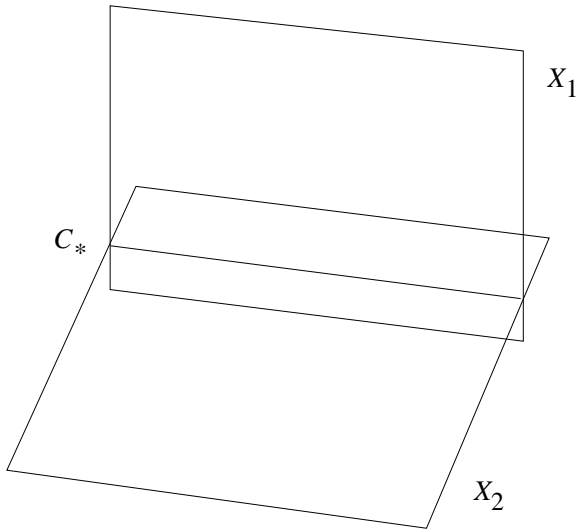,height=7cm}}
   \caption[hpso16]{The stable degeneration of $X$}
\label{fig:degen}
\end{figure}

Now we turn to the stable degeneration limit introduced
in~\cite{FMW,AM1}. In the limit where the heterotic K3 manifold $Y$ is
taken to the large, the $\bbP^1$ fibre of the F-theory $\bbF_n$ base
degenerates into a pair of intersecting $\bbP^1$ curves. This can be
viewed as one $S^1$ cycle in the two-sphere $\bbP^1$ pinching to a
point. The F-theory base then degenerates into a pair of Hirzebruch
surfaces, $\bbF_{n,1}$ and $\bbF_{n,2}$, intersecting over a $\bbP^1$
curve $C_*$. This is a section in the class $C_*=C_0+nf$ on each
$\bbF_n$ surface. The full F-theory threefold $X$ is a degeneration
into a pair of threefolds $X_1$ and $X_2$ intersecting in a K3
surface, which is the elliptic fibration over $C_*$. This is shown in
Figure~\ref{fig:degen}. 
The intersection K3 surface can then be identified with the heterotic
K3 surface $Y$. Roughly, one can identify each threefold with one
$E_8$ group of the heterotic string. Because of the degeneration, the
elliptically fibred threefolds $\pi_1:X_1\to\bbF_{n,1}$ and
$\pi_2:X_2\to\bbF_{n,2}$ are no longer Calabi--Yau. Instead, the
canonical bundles pick up a contribution from the pull-back of $C_*$,
so that, as classes, $K_{X_1}=\pi_1^*(C_*)$ and
$K_{X_2}=\pi_2^*(C_*)$. Consequently, the class $L$ of the co-normal
bundle $\mathcal{L}$ on each threefold is now given by
$L=-(K_{\bbF_{n,1}}+C_*)$, so that 
\begin{equation}
   L = C_0 + 2f
\end{equation}
on each of $\bbF_{n,1}$ and $\bbF_{n,1}$. As before, in the
Weierstrass model of $X_1$ and $X_2$, the polynomials $a$, $b$
and $\delta$ are still sections of $\mathcal{L}^4$, $\mathcal{L}^6$
and $\mathcal{L}^{12}$ respectively.

\subsection{The $\SO(16)\times[\SU(2)\times E_7]$ model}
\label{sec:Fmodel}

We now discuss the particular F-theory geometry corresponding to the
perturbative $\SO(16)\times[\SU(2)\times E_7]$ model. By restricting
ourselves to the Weierstrass model~\eqref{weierst} we naturally
describe a different $\bbZ_2$-orbifold limit of the heterotic
K3 from $T^4/\bbZ_2$. However, it is only the local geometry near each
$A_1$ singularity which encodes the information relevant to
the M-theory rules and so this model is quite sufficient. In fact, it
is relatively easy to generalize the discussion to get the full global
$T^4/\bbZ_2$ model if required.  

Recall that the $E_8$ gauge bundles in the heterotic limit had
discrete $\bbZ_2$ holonomy. The F-theory duals of such models have
been described explicitly by Aspinwall and Morrison~\cite{AM2}. The
following discussion is the direct analogue of their $\bbZ_3$
example. 

Requiring that we have $\bbZ_2$ holonomy restricts the polynomials
$a$, $b$ in the Weierstrass model to have a particular form given by 
\begin{equation}
\begin{aligned}
   a &= a_4 - \frac{1}{3}a_2^2, \\
   b &= \frac{1}{27}a_2 \left( 2a_2^2 - 9a_4 \right), \\
   \delta &= a_4^2 \left( 4a_4 - a_2^2 \right)
\end{aligned}
\label{fracinst}
\end{equation}
where $a_i$ is a section of $\mathcal{L}^i$. Recall that the two
possible preserved gauge groups for a $\bbZ_2$-bundle in $E_8$ are
$\SO(16)$ and $\SU(2)\times E_7$. From the Table~\ref{tab:sing} we
expect these to correspond to Kodaira fibres of type $\KI^*_4$ and
$(\KI_2+\KIII^*)$ respectively.

\begin{figure}[htbp]
   \centerline{\epsfig{file=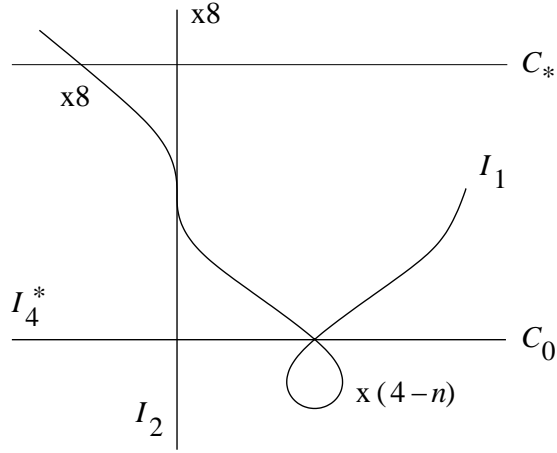,height=6cm}}
   \caption[hpso16]{${\mathbb{F}_{n,1}}$ with a perturbative gauge
      group $SO(16)$ and instantons with $\mathbb{Z}_2$ holonomy}  
\label{hpso16}
\end{figure}

Let us first consider the threefold $X_1$ with gauge group
$\SO(16)$. Following the usual prescription the unbroken perturbative
gauge group is given by singular fibers over the exceptional divisor
$C_0$ on $\bbF_{n,1}$. Let $c_0=0$ define this divisor. For $\SO(16)$
we need $\KI^*_4$ fibres, so, from Table~\ref{tab:sing}, we
see that the polynomials $a, b$ and $\delta$  vanish to orders 
$2,3$ and $10$ on $C_0$.  This implies that the polynomials $a_2$ and
$a_4$ have to be of the form  
\begin{equation}
   a_2 = c_0 g , \qquad
   a_4 = c_0^4 h ,
\label{eq:a24so}
\end{equation}
where $g=0$ and $h=0$ define divisors in the classes $C_0+4f$ and $8f$
respectively. Generically, curves in the class $8f$ split into eight
distinct fibres, so the polynomial factors into $h=f_1\ldots
f_8$. The discriminant curve is then given by
\begin{equation}
   \delta = c_0^{10} f_1^2 \dots f_8^2 
      \left(4c_0^2f_1\dots f_8-g^2\right) ,
\label{eq:dso}
\end{equation}
together with 
\begin{equation}
\begin{aligned}
   a &= c_0^2 \left(c_0^2f_1\dots f_8 - \frac{1}{3} g^2\right) , \\ 
   b &= \frac{1}{27} c_0^3 g \left(2g^2 - 9c_0^2f_1\dots f_8\right) .
\end{aligned}
\label{eq:abso}
\end{equation}
From the factors in $\delta$, again comparing with
Table~\ref{tab:sing}, we see that there are eight curves $f_i=0$ of 
$\KI_2$ fibres, giving eight $\SU(2)$ groups in addition to the
$\SO(16)$ factor. The factor $k\equiv 4c_0^2f_1\dots f_8-g^2=0$, gives a
divisor in $2C_0+8f$ which generically gives a single curve with
$\KI_1$ fibres and no additional gauge groups. It is easy to show that
this curve has $(4-n)$ double intersections with $C_0$. It also has a
single tangential intersection with each $f_i=0$ and eight transversal
intersection with $C_*$. This is illustrated in Figure~\ref{hpso16}. 

\begin{figure}[htbp]
   \centerline{\epsfig{file=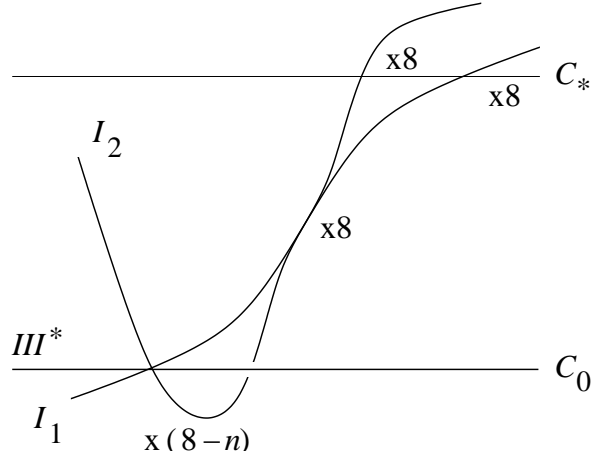,height=6cm}}
   \caption[hpe7]{${\mathbb{F}_{n,2}}$ with perturbative gauge group
       $E_7$ and instantons with $\mathbb{Z}_2$ holonomy} 
\label{hpe7}
\end{figure}

Now turn to the second threefold $X_2$ with perturbative gauge group
$\SU(2)\times E_7$. This implies that there are singular fibres of type
$\KIII^*$, with $a$, $b$ and $\delta$ vanishing to order $3$, at least
$5$ and $9$ respectively over the exceptional divisor $C_0$. This implies
\begin{equation}
   a_2 = c_0^2 g , \qquad  a_4 = c_0^3 h,
\end{equation}
where $g=0$ and $h=0$ define divisors in the classes $4f$  and
$(C_0+8f)$ respectively. In analogy to the previous case,  $g$
factorises into $g=f_1\dots f_4$. The discriminant curve is given by
\begin{equation}
   \delta = c_0^9 h^2 \left(4h - c_0g^2\right) ,
\label{eq:de7}
\end{equation}
while
\begin{equation}
\begin{aligned}
   a &= c_0^3 \left(h - \frac{1}{3} c_0 g^2\right) \\
   b &= \frac{1}{27} c_0^5 g \left(2c_0g^2 - 9h\right) .
\end{aligned}
\end{equation}
We see that, in addition to the $\KIII^*$ fibres over $c_0=0$ giving
an unbroken $E_7$ gauge group, we have a curve $h=0$ of $\KI_2$
fibres, giving an additional $\SU(2)$ factor as expected. The
remaining factor in $\delta$, gives a divisor $k\equiv 4h-c_0g^2=0$ in
the class $C_0+8f$, generically giving a single curve of $\KI_1$
fibres, and no additional gauge factors. It is easy to show that there
are $8-n$ points on $C_0$ where the curve $h=0$ of $\KI_2$ fibres and
the curve $k=0$ of $\KI_1$ both intersect transversally. In addition,
these curves also intersect tangentially at eight points away from
$C_0$. Finally, both curves also intersect $C_*$ eight times
transversally. This is illustrated in Figure~\ref{hpe7}.  

\begin{figure}[htbp]
   \centerline{\epsfig{file=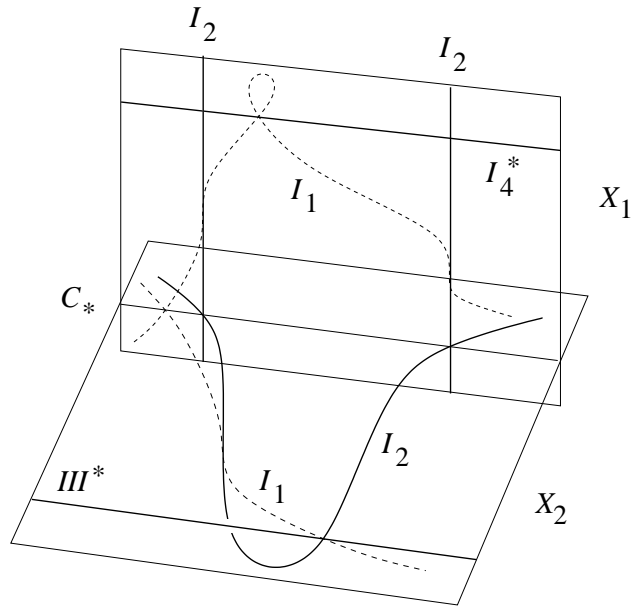,height=8cm}}
   \caption[hpe7]{The full degenerate space $\bbF_{n,1}\vee\bbF_{n,2}$} 
\label{fig:X}
\end{figure}

To reconstruct the complete degenerated threefold, we have to glue the
bases of the two threefold $X_1$ and $X_2$ together along $C_*$. To be
consistent, the singular fibres over $C_*$ must be the same on each of
$X_1$ and $X_2$. Recall that for the $\SO(16)$ factor $X_1$ had a
single curve of $\KI_1$ fibres intersecting $C_*$ eight times and
eight curves of $\KI_2$ fibres intersecting $C_*$ once. These
intersections must then match those on the $\SU(2)\times E_7$ factor
$X_2$ which also had a single curve of $\KI_1$ fibres intersecting $C_*$
eight times together with a \textit{single} curve of $\KI_2$ fibres
intersecting $C_*$ eight times. The full degenerated threefold is
shown in Figure~\ref{fig:X}. Note that $C_*$ has eight $\KI_1$ fibres
and eight $\KI_2$ fibres and so is indeed a K3 surface, reproducing the
heterotic K3 surface $Y$. The $\KI_2$ fibres give eight
$A_1$ singularities. Clearly, although locally near the
orbifold points we have the same geometry and perturbative bundles as
the $T^4/\bbZ_2$ model, globally we have a different limit of the K3
manifold since we have only eight fixed points and not sixteen. 

\subsection{M-theory limit}
\label{sec:Mlimit}

Having constructed the F-theory model we can now take the M-theory
limit to compare with the rules given in~\cite{FLO,KSTY}. Recall that
the stable degeneration corresponds to shrinking one of the $S^1$
cycles in the $\bbP^1$ fibre of $\bbF_n$ to a point, to form a pair of
intersecting $\bbP^1$ curves. Following~\cite{A} (see
also~\cite{AM4}), taking the M-theory limit requires shrinking a whole
family of $S^1$ cycles to points, so that the sphere $\bbP^1$ becomes
a one-dimensional interval. This reduces the threefold $X$ to a
five-dimensional manifold and which can be viewed as the M-theory
compact space $S^1/\bbZ_2\times Y$.

The five-dimensional manifold can be still represented by the diagram,
Figure~\ref{fig:X}, though now the $\bbP^1$ fibres of $\bbF_{n,1}$ and
$\bbF_{n,2}$ must be viewed as forming a single $S^1/\bbZ_2$ interval
bounded by the $C_0$ sections which represent the fixed
ten-planes. The manifold $Y$ is fixed by the K3 surface over
$C_*$. The $A_1$ orbifold singularities are described by
the curve of $\KI_2$ fibres. Note that on $X_1$ these intersect $C_0$
transversally and the space looks something like a product $Y\times
S^1/\bbZ_2$, with the lines of $\KI_2$ fibres describing the fixed
seven-planes. However, on $X_2$, the curve of $\KI_2$ fibres has a
component parallel to the $C_0$ and so the fixed-seven planes are
distorting.  From this it is clear that the geometry of the full
M-theory space is not simply a product. This supports the similar
conclusion reached above arguing directly from M-theory. 

What gauge and matter fields do we find present in the M-theory limit?
First consider the gauge groups. We have $\SO(16)$ and $E_7$ factors
from singular fibres over the $C_0$ sections for $X_1$ and $X_2$
respectively. On $X_1$ it appears we have eight distinct $\SU(2)$
factors, one for each $A_1$ singularity on $Y$. These correspond to
the fixed seven-planes of the M-theory $T^4/\bbZ_2\times S^1/\bbZ_2$
manifold. However, from the intersection over $C_*$, we see that each
of these connects to the single curve of $\SU(2)$ singularities on
$X_2$. Thus over the whole (degenerate) space $X$, there is only a
\textit{single} curve of $\SU(2)$ singularities, which is the source
of the $\SU(2)^*$ factor in $\SO(16)\times[\SU(2)^*\times E_7]$. In
other words, we see \textit{directly} that the $\SU(2)$ factors on the
the fixed seven-planes must be identified with the perturbative $\SU(2)$,
justifying the arguments in~\cite{FLO,KSTY}. 

The matter appears as follows. First, there are the usual perturbative
hypermultiplets $(\rep{128},\rep{1},\rep{1})$ and
$(\rep{1},\rep{2},\rep{56})$ in $\SO(16)\times[\SU(2)\times E_7]$. In
F-theory these correspond to the possibility of blowing-up the
singularities and Higgsing the preserved gauge
group~\cite{Ftheory}. However, in addition, we get extra matter when
different parts of the discriminant curve
intersect~\cite{Ftheory,BSV,KV,AKM}. In particular, the intersection of
the curve of $\KI_2$ fibres and the $\KIV^*$ fibres over $C_0$ in
$X_1$, lead to sixteen half hypermultiplets in fundamental
representations  $\frac{1}{2}(\rep{16},\rep{2},\rep{1})$. This is
precisely the twisted matter of the perturbative heterotic string. It
is easy to see how this can be charged under gauge groups coming from
\textit{both} of the fixed ten-planes, because, in the distorted
M-theory geometry, the $\SU(2)$ gauge group is seen to ``stretch''
across the $S^1/\bbZ_2$ orbifold to intersect the $\SO(16)$ fixed
plane. Again, we find an F-theory justification for the arguments
in~\cite{FLO,KSTY} as to how the twisted matter appears. 

Recall that, while locally the F-theory model gave a heterotic K3
surface with $A_1$ singularities, globally the surface was not
$T^4/\bbZ_2$. This is reflected by the fact that there were only eight
$A_1$ singularities and, in addition, a curve of $\KI_1$ fibres on
$X$. On $X_1$ this curve has $(4-n)$ double intersections with
$C_0$. Following~\cite{AM1} these are interpreted as ordinary
pointlike instantons on $Y$. As a result, there are an additional
$(4-n)$ tensor multiplets in the spectrum, parametrising the
``Coulomb'' branch describing the motion of the instantons into the
$S^1/\bbZ_2$ bulk as M5-branes~\cite{GH,SW}. Similarly, there are
$(8-n)$ mutual intersections of the $\KI_1$, $\KI_2$ and $C_0$ curves
in $X_2$ leading to a further $(8-n)$ pointlike-instanton tensor
multiplets. Note that this provides a check that the instanton
charge of the fractional instantons at the $A_1$ singularities in the
model is as expected. Recall that each $\bbF_{n}$ plane carries a
total instanton number of $12-n$~\cite{Ftheory,AM1}. Each ordinary
pointlike instanton carries charge one. Thus the eight fractional
instantons on $X_1$ must also each carry charge one. On $X_2$ however,
each of the eight fractional instantons must carry charge $1/2$. This
matches exactly the expected charges for $\SO(16)$ and $\SU(2)\times
E_7$ perturbative bundles discussed above. 

Let us end this section by briefly discussing how the F-theory model
should be modified so that one realises the exact global dual of the
$T^4/\bbZ_2$ compactification rather than simply the local behaviour
near the $A_1$ singularities. The essential point is that $T^4/\bbZ_2$
cannot be realized directly as a conventional Weierstrass model of
the form~\eqref{weierst} with $\KI_2$ fibres. Instead one takes a
model with four $D_4$ singularities, giving gauge group $\SO(8)^4$ and
then blows up one cycle in each fiber to Higgs the $\SO(8)$
group to $\SU(2)^4$, giving four $A_1$ singularities per fibre. To
realise the F-theory dual, we fix $n=4$ on the $E_7$  half-plane and
$n=-4$ on the $\SO(16)$ plane. To reproduce the four $D_4$
singularities on $C_*$ we have to restrict to the case where the eight
$\KI_1$ fibres and eight $\KI_2$ fibres on $C_*$ come together in four
sets of two $\KI_1$ and two $\KI_2$. This restricts the form of the
discriminant curves~\eqref{eq:dso} and~\eqref{eq:de7}. In particular,
on the $\SO(16)$ plane the function $h$ in~\eqref{eq:a24so} must
factor as  $h=f_1^2f_2^2f_3^2f_4^2$ to give four curves of
$\KI_0^*$ fibres with $D_4$ singularities.


\section{Other cases and discussion}
\label{sec:others}

Let us end by mentioning how this analysis can be extended to other
orbifold compactifications, starting with other examples on
$T^4/\bbZ_2$. The F-theory analysis justified local rules for gauge
groups and matter at the $A_1$ singularities. In particular, for a
charge $1/2$ fractional instanton leaving the gauge group $\SU(2)\times
E_7$, we saw that there was an $\SU(2)$ gauge group on the fixed
seven-plane which is identified with the perturbative $\SU(2)$. No
additional matter appeared at the intersection of the seven- and
the ten-plane. For a unit charge fractional instanton leaving gauge
group $\SO(16)$, there was again an $\SU(2)$ gauge group on the
seven-plane, and now additional hypermultiplets at the intersection
transforming as $\frac{1}{2}(\rep{16},\rep{2})$ under
$\SO(16)\times\SU(2)$. 

In~\cite{FLO} further rules have been developed for other types
of instanton at $A_1$ singularities. In particular, one very obvious
case to consider is the conventional perturbative heterotic background
with the standard embedding. This has charge $3/2$ fractional
instantons on one ten-plane and none on the other, leaving a gauge
group $E_8\times[\SU(2)\times E_7]$. Notably, unlike the example
above, this background has no perturbative states charged under gauge
groups from different ten-planes. In turns out that, as shown by
Aspinwall and Donagi~\cite{AD} the F-theory dual of the standard
embedding is particularly subtle. In particular, it is crucial that
one considers the role of the Ramond--Ramond (RR) fields in order to
distinguish it from other duals. The effect is that for non-zero RR
backgrounds less gauge symmetry is preserved than might immediately
appear from the geometry. Nonetheless, one would expect that a similar
analysis of the M-theory limit as above is possible. A further
generalization in~\cite{FLO,KSTY}, are models with
$E_8\times[\SU(2)\times E_7]$ or $E_8\times E_8$ perturbative symmetry
and additional $U(1)$ non-perturbative factors. In general, $U(1)$
factors are hard to identify in F-theory models, though are still
encoded in the geometry as discussed for instance in~\cite{G}. 

One model that can be easily analysed is that with $\SO(8)^8$ gauge
group considered in~\cite{GM}. This was argued to be dual to F-theory
on $T^6/(\bbZ_2\times\bbZ_2)$, with a base manifold $T^2/\bbZ_2\times
T^2/\bbZ_2$ which is a singular limit of
$\bbF_0=\bbP^1\times\bbP^1$. Following~\cite{AM2}, in this case we
expect that the heterotic background has $\bbZ_2\times\bbZ_2$
fractional instantons at the $A_1$ singularities, preserving a
perturbative $\SO(8)^2$ for each $E_8$ factor. The Weierstrass model
then takes the form 
\begin{equation}
\begin{aligned}
   a &= \frac{1}{3}(b_2c_2 - b_2^2 - c_2^2) , \\
   b &= -\frac{1}{27}(b_2 + c_2)(b_2 - 2c_2)(2b_2 - c_2) , \\
   \delta &= - b_2^2c_2^2(b_2 - c_2)^2 ,
\end{aligned}
\end{equation}
where $b_2$ and $c_2$ are sections of $\mathcal{L}^2$. Taking the
stable degeneration, one gets the correct gauge group by taking 
\begin{equation}
   b_2 = c_1 c_2 f_1 \dots f_4
\end{equation}
one each of $\bbF_{0,1}$ and $\bbF_{0,2}$. The functions $c_1$ and
$c_2$ define two different sections in the divisor class of the first
$\bbP^1$ factor in $\bbF_0=\bbP^1\times\bbP^1$, and $f_1$ to $f_4$
give four different sections in the class of the second factor.  The
two $c_1=0$ and $c_2=0$ curves intersect the four $f_i=0$ curves
transversally. From the discriminant curve, each function defines a
curve of $\KI_0^*$ fibres giving an $\SO(8)$ factor. Gluing $X_1$ and
$X_2$ together, we must identify the two sets of $f_i=0$ curves and so
the full gauge group becomes $\SO(8)^8$.  Four $\SO(8)$ factors are
perturbative and four non-perturbative. As pointed out in~\cite{GM},
there is a symmetry exchanging perturbative and non-perturbative
factors, essentially by exchanging the role of the two $\bbP^1$
factors in each $\bbF_0$. Note that as stands this is not quite the
required model. As in the discussion at the end of the last section,
the K3 manifold over the $C_*$ intersection is not $T^4/\bbZ_2$ but a
more singular space with four $D_4$ singularities. To obtain
$T^4/\bbZ_2$ we must blow up one curve in each singular fibre, thus
Higgsing $\SO(8)$ to $\SU(2)^4$. In general this blows up the
corresponding fibres in the $f_i=0$ curves so the full symmetry is
actually $\SU(2)^{16}\times\SO(8)^4$. To preserve the symmetry between 
perturbative and non-perturbative groups, one would also blow up one
cycle in the fibres above each of the $c_1=0$ and $c_2=0$ sections so
that the final gauge symmetry becomes  $(SU(2))^{32}$. 

The other obvious class of generalisations is to other K3
orbifolds. Again~\cite{FLO,KSTY} give rules for
instantons at other types of singularity. The most straightforward
generalisation is the $\bbZ_3$ version of the $\bbZ_2$ model
considered above. The orbifold $T^4/\bbZ_3$ has nine $A_2$
singularities and  there are two types of $\bbZ_3$-holonomy
bundles, preserving either $\SU(9)$ or $\SU(3)\times E_6$. The
perturbative model with both groups $\SU(9)\times[\SU(3)\times E_6]$ is
the analogue of the $\bbZ_2$ example considered above. This example
was explicitly worked out in~\cite{AM2}. One finds a Weierstrass model
with  
\begin{equation}
\begin{aligned}
   a &= a_1\left(\frac{1}{2}a_3-\frac{1}{48}a_1^3\right) , \\
   b &= \frac{1}{4}a_3^2 + \frac{1}{864}a_1^6 
         - \frac{1}{24}a_1^3a_3 , \\
   \delta &=\frac{1}{16}a_3^3(27a_3-a_1^3) .
\end{aligned}   
\end{equation}
where $a_i$ is a section of $\mathcal{L}^i$. In the stable
degeneration limit, on the $\SU(9)$ threefold one takes
\begin{equation}
   a_1 = g , \qquad
   a_3 = c_0^3 f_1 \dots f_6 ,
\end{equation}
where $c_0$ vanishes on $C_0$, the $f_i$ vanish on distinct fibres and
$g=0$ is in the class $C_0+2f$. This gives a curve of $\KI_9$ fibres
with $\SU(9)$ on $C_0$, six curves $f_i=0$ of $\KI_3$ fibres with
$\SU(3)$ and a single curve of $\KI_1$ fibres. On the $\SU(3)\times
E_6$ threefold one takes 
\begin{equation}
   a_1 = c_0 f_1 f_2 , \qquad
   a_3 = c_0^2 h ,
\end{equation}
where the $f_i$ vanish on distinct fibres and $h=0$ is in the class
$C_0+6f$. This gives a curve of $\KIV^*$ fibres with $E_6$ on $C_0$, a
single curve of $\KI_3$ fibres with $\SU(3)$ on $h=0$ and a single
curve of $\KI_1$ fibres. Again, gluing $X_1$ and $X_2$ means that the
$\SU(3)$ factors are identified as a single curve so the final gauge
symmetry becomes $\SU(9)\times[\SU(3)^*\times E_6]$ in direct
analogy with the $\bbZ_2$ example. There are related models with
$\bbZ_4$ and $\bbZ_6$ symmetry which it should be possible to analyse
in an analogous fashion.  

In summary, we have shown that known results for the F-theory
duals of different Ho\v{r}ava--Witten orbifold compactifications
provide a good explanation of some of the more counter-intuitive rules
implied by anomaly cancellation in the M-theory model. In particular,
it becomes clear why in some cases gauge groups on the fixed
seven-planes should be identified with perturbative gauge groups on
the fixed ten-planes. Also, this provides additional evidence that the
actual M-theory geometry is not simply a product. One interesting
extension would be to explore this geometry further directly in the
M-theory model. It also appears to be possible to extend the analysis
to various other orbifold models and perhaps use this approach to
analyse M-theory orbifold compactifications to four dimensions.


\subsection*{Acknowledgements}

Both authors are supported in part by PPARC through the grant SPG
$\#$613. DW also thanks the Royal Society for support and the Isaac
Newton Institute at the University of Cambridge for hospitality during
the completion of this manuscript. The work of MM  was also supported by the University of Halle
and a Marie Curie Postgraduate Fellowship by the EU. MM thanks the
Physics Department of Queen Mary College, University of London, for hospitality while the work
presented here was done.  



\end{document}